\documentclass[%
 reprint,
 amsmath,amssymb,
 aps, physrev,
]{revtex4-2}

\usepackage{graphicx}
\usepackage{dcolumn}
\usepackage{bm}

\usepackage[textsize=footnotesize]{todonotes}
\usepackage{hyperref}

\begin{document}

\title{{Steering Non-Equilibrium Molecular Dynamics in Optical Cavities}}
\author{Mingxuan Xiao$^{1}$}
\author{Wei Wang$^{1}$}
\author{Wenjing Liu$^{1,3}$}
\author{Zheng Li$^{1,3}$}
\author{Shui-Jing Tang$^{2}$}
\author{Yun-Feng Xiao$^{1,3}$}

\affiliation{
$^1$State Key Laboratory for Mesoscopic Physics, Frontiers Science Center for Nano-optoelectronics, New Cornerstone Science Laboratory, School of Physics, Peking University, 100871, Beijing, China\\
$^2$National Biomedical Imaging Center, College of Future Technology, Peking University, Beijing 100871, China\\
$^3$Collaborative Innovation Center of Extreme Optics, Shanxi University, Taiyuan 030006, China}







\date{\today}

\begin{abstract}
Optical resonators have shown outstanding abilities to tailor chemical landscapes through enhanced light-matter interaction between confined optical modes and molecule vibrations.
We propose a theoretical model to study cooperative vibrational strong coupling in an open quantum system.
The non-equilibrium stochastic molecular dynamics in an optical cavity with an auxiliary ensemble is investigated. It shows that coupling with a cavity mode introduces an additional colored noise and a negative feedback, both of which enable control over thermalization rates (i.e. the lifetime of excitation) of reactive molecules.
Our work offers a pathway to steer stability of chemical bonds for chemical reactivity under cooperative vibrational strong coupling.
\end{abstract}

\maketitle


Light-matter interactions between optical cavities and molecular vibrational transitions have sparked growing interest. 
This is attributed to the achievement of collective strong coupling, which enables energy to exchange coherently between the optical field and collective molecular vibrations before dissipation occurs \cite{shalabney2015coherent}.
This interaction, facilitated by resonant dipolar coupling and strengthened by optical field confinement within cavities \cite{vahala2003optical}, 
has proven effective in altering the rates of chemical reactions \cite{thomas2016ground} 
and selectivity \cite{thomas2019tilting} without external laser excitation.
In the collective and cooperative coupling regime \cite{lather2019cavity,lather2020improving,thomas2020ground,lather2022cavity,wang2021roadmap,nagarajan2021chemistry,mandal2023theoretical,schutz2020ensemble}
each molecular unit is essentially weakly coupled and visible Rabi splitting is achieved by large numbers of molecules.
Pioneering experiments have demonstrated modified reactivity in various reactions,
such as Prins cyclization \cite{hirai2020modulation},
charge transfer \cite{pang2020role},
and enzyme hydrolysis \cite{vergauwe2019modification}.
By cavity tuning \cite{ahn2023modification}, 
the kinetic isotope effect \cite{lather2019cavity}, 
and the dependence on Rabi splitting \cite{thomas2020ground}, 
the observed modification on reactivity is confirmed as a genuine cavity effect.
Analytical and computational efforts have been aimed at revealing the mechanism behind cavity quantum electrodynamics (cQED) chemistry.
The investigated pathways range from equilibrium modification in transition state theory \cite{galego2019cavity,li2020origin,campos2020polaritonic},
dynamical caging effect \cite{li2021collective,li2021cavity,mandal2022theory},
dark state delocalization \cite{du2022catalysis}, acceleration in vibration energy relaxation \cite{li2021collective}
and redistribution \cite{schafer2022shining},
and deviation of vibrational populations from Boltzmann statistics in thermal equilibrium \cite{ahn2023modification}. 
However, most theories have primarily concentrated on closed quantum systems, often ignoring the effects of noise and decay owing to interaction with external reservoirs or addressing thermal equilibrium in open quantum systems. Consequently, the mechanisms and processes in which thermal equilibrium is reached remain largely unexplored.

The non-equilibrium dynamics of open quantum systems offers us a tool to track whether and how such systems reach thermal equilibrium with an exterior reservoir of finite temperature. 
In molecule physics, it has been demonstrated that a non-Boltzmann equilibrium can be established when an optical cavity is introduced into the system \cite{ahn2023modification},
and the qualitative analysis of frequency dependence shed light on the intriguing question of how such a non-canonical distribution is reached.
Meanwhile, the dynamics in open quantum systems features many properties reminiscent of those in cQED chemistry.
Stochastic resonance (SR), for example, occurs when the frequency of an external periodic driving is tuned to the thermal hopping rate in a nonlinear system \cite{gammaitoni1998stochastic,zhou1990escape,gammaitoni1995stochastic,lofstedt1994quantum,grifoni1996coherent}.
In SR, the response to periodic driving could be drastically enhanced, providing a potential pathway to amplify the weak cavity electromagnetic force to produce observable effects in molecular systems.
In a non-Markovian process, molecules have been shown to reach thermal equilibrium much faster than in a Markovian environment \cite{ceriotti2009langevin}. This finding opens avenues for investigating cQED chemistry through the accelerated stabilization of reactants or products.

\begin{figure*}
    \centering
    \includegraphics[width=16 cm]{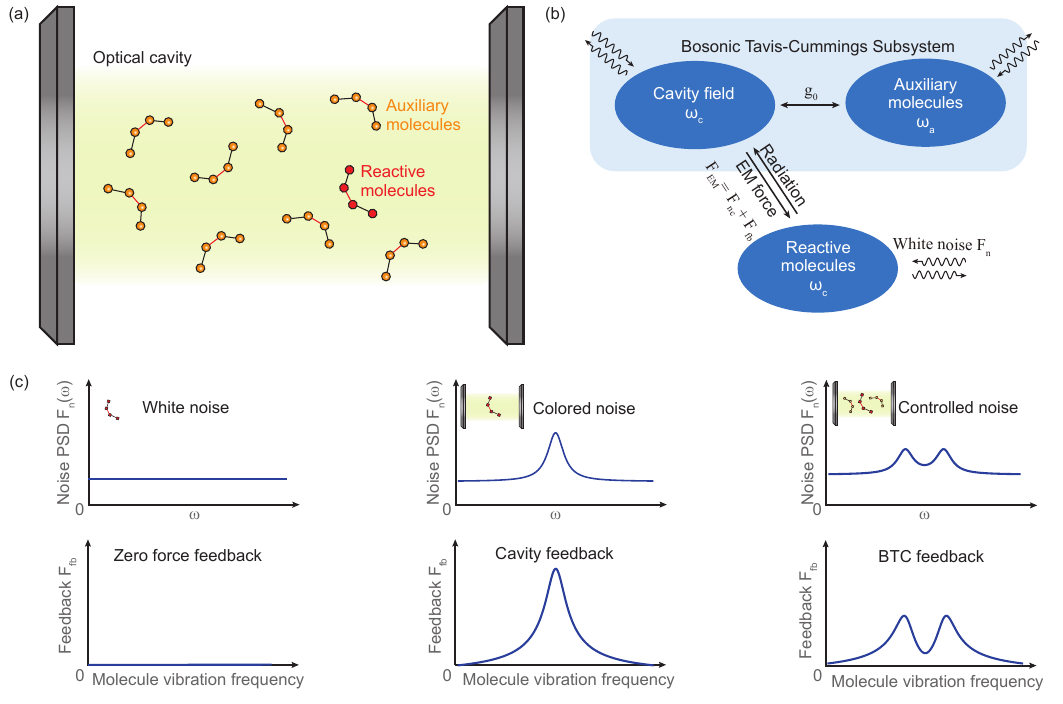}
    \caption{(a) Schematics of cooperative vibrational strong coupling. The reactive molecules of interest are placed in an optical resonator, together with an auxiliary molecule ensemble. The cavity is tuned to couple with a certain vibrational mode of the reactive molecules, and an auxiliary ensemble with vibrational mode close to resonance is also introduced to modulate the spectrum. Noise and decay are introduced into the coupled molecule-cavity system to reflect open quantum system effects.
    (b) Schematics of the model Hamiltonian. The cavity mode and the auxiliary ensemble
    form a bosonic Tavis-Cummings (BTC) subsystem, which couples to the reactive subsystem as a background. Both subsystems are made open by coupling to exterior environments.
    (c) Cavity effects introduced by the optical cavity and controlled by the auxiliary ensemble. The introduction of cavity brings in a non-Markovian noise, adding to the white noise in the vibrations of bare molecules. The power spectrum density (PSD) of the noise could be tuned by an auxiliary ensemble. A feedback response generated by backaction from the optical mode to the reactive molecules is effectively turned on via the molecule-cavity interaction, and the dependence of its amplitude on molecule vibration frequency is shown.
    }
    \label{fig:1}
\end{figure*}

Here, we investigate the non-equilibrium dynamics of an open system consisting of molecules coupled with a cavity, 
and show that this interaction effectively results in an extra non-Markovian noise and a coherent feedback action to the reactive system, 
the resonant properties of the two cavity effects, i.e., the non-Markovian noise and the feedback action, are adjustable by an auxiliary molecule ensemble.
We show that the added colored noise results from the memory effect of the optical polariton modes 
and the sign of the coherent feedback is always negative.
We have also simulated the stochastic dynamics, demonstrating shorter thermalization time.
This accelerated thermalization allows molecules to decay to lower energies in a shorter time,
providing a pathway to modify chemical reactivity by stabilizing reactants, intermediates, or products.
Furthermore, while the setup we propose here works in an experimentally achievable coupling strength,
we also show how both cavity effects scale down with decreasing single-molecule coupling strength.
This indicates the existence of an amplification effect not hereby included if the proposed non-Markovian environment and negative feedback is to produce catalysis comparable to that observed experimentally in typical collective and cooperative vibrational strong-coupling regimes of cQED chemistry.

The system we investigate consists of a swarm of reactive molecules,
a single cavity mode of an optical resonator, and
an auxiliary molecule ensemble (Fig.~\ref{fig:1}(a)) with coupling strength in the experimentally achievable weak coupling regime \cite{chikkaraddy2016single}.
We model the system with a Hamiltonian of three parts (Fig.~\ref{fig:1}(b)),
the reactive molecules moving on a potential energy surface,
an ensemble of harmonic oscillators, and the cavity electromagnetic field,
each subsystem is made open by including noise and decay.
The latter two parts of the Hamiltonian function as the background for the reactive subsystem, which consists of multiple bosonic modes that interact with a single cavity mode, and are denoted as the Tavis-Cummings (BTC) bosonic subsystem owing to their equivalence \cite{tavis1968exact}.
The full Hamiltonian (see Supplemental Materials (SM) for complete derivation) \cite{suppl}
for the coupled system reads:
\begin{subequations}
\label{eq:aux}
\begin{equation}
H=H_{\mathrm{reac}}+H_{\mathrm{BTC}}+H_{\mathrm{coupl}}
\end{equation}
\begin{equation}
H_{\mathrm{reac}}=\sum_j \frac{\textbf{P}_j^2}{2M_j}+E(\textbf{R})
\end{equation}
\begin{eqnarray}
H_{\mathrm{BTC}}=&&\hbar \omega_{\mathrm{c}} a^\dagger a +\hbar\sum_{k=1}^{N}  [\omega_{a} b_{k}^\dagger b_{k}+g_{k}(a^\dagger b_{k}+a b_{k}^\dagger)]
\end{eqnarray}
\begin{equation}
H_{\mathrm{coupl}}=\sqrt{\frac{\hbar\omega_{\mathrm{c}}}{2\epsilon_0V}}\hat{e}\cdot\mathbf{\mu}(\textbf{R})(a^\dagger+a)
\end{equation}
\end{subequations}
where \(a^\dagger \) and \(a\) are the creation and annihilation operators of the optical mode of the cavity with angular frequency \( \omega_{\mathrm{c}} \),
\(b_k^\dagger\) and \(b_k\) are those of the resonating
mode of the $k$-th auxiliary molecule whose angular frequency is denoted by \(\omega_{\mathrm{a}}\),
and \(g_k\) is the single-molecule coupling strength
between an auxiliary molecule and the optical field.
The two terms in \(H_{\mathrm{reac}}\) represent the kinetic energy and the potential energy of the nuclei configuration, respectively, 
and \( H_{\mathrm{coupl}} \) depicts the dipolar coupling between the two subsystems,
with \(\mathbf{\mu}\) being the dipole moment of the configuration and \(V\) the mode volume of the optical field.
In the following, we focus on the weak coupling regime between the reactive subsystem and the BTC subsystem, while the BTC subsystem is in its own collective strong coupling regime.
The coupling between the reactive molecules and the BTC subsystem is below the loss of both the cavity and the auxiliary ensemble,
thus the reactive subsystem does not split the BTC modes distinguishably.

We first examine the properties of the BTC subsystem, which serves as a modified background for the reactive subsystem. 
The ensemble interacts collectively with the optical mode,
giving two bright states separated by a Rabi splitting of \(\sqrt{N}g_{\mathrm{a}}\), where \(g_{\mathrm{a}}\) is the root mean square (RMS) value of all \(g_k\).
Once the splitting exceeds both the cavity decay \(\kappa\) and molecular loss \(\gamma_{\mathrm{a}}\),
the BTC subsystem enters the collective strong coupling regime manifested
by the peaked structure in the transmission spectrum (see SM for transmission from the input-output formalism) \cite{suppl}.
The other \(N-1\) dark states are not visualized in the transmission spectrum
because of the absence of photonic components in the dark states, which could be shown by diagonalizing the Hamiltonian \cite{suppl}.

We now focus on the dynamics of the two interacting subsystems.
The BTC subsystem has two polaritonic eigenmodes,
with creation operator \(c_{\mathrm{l}}^\dagger\) for the lower polariton branch 
and \(c_{\mathrm{u}}^\dagger\) for the upper branch,
the equations of motion with noise and decay read:
\begin{subequations}
\label{eq:eom}
\begin{eqnarray}
\label{eq:eom_btc}
\frac{\mathrm{d} c_k}{\mathrm{d} t}=&&-i \omega_k c_k-\frac{1}{2}\Gamma_k c_k+\sqrt{\Gamma_k\bar{n}^{\mathrm{th}}_{k}}f_{k}(t)\nonumber\\
&&-\frac{i}{\hbar}\xi_{k}\sqrt{\frac{\hbar \omega_{\mathrm{c}}}{2\epsilon_0 V}}\hat{e}\cdot \mathbf{\mu}(\mathbf{R})\,\,\,(k=\mathrm{l,u})
\end{eqnarray}
\begin{eqnarray}
M\frac{\mathrm{d}^2 \mathbf{R}}{\mathrm{d} t^2}=&&-\nabla \left[ E(\mathbf{R})\right]- M \alpha \frac{\mathrm{d} \mathbf{R}}{\mathrm{d} t}+\sqrt{2M\alpha k_{\mathrm{B}}T}F_0(t)\nonumber\\
&&-\sqrt{\frac{\hbar \omega_{\mathrm{c}}}{2\epsilon_0 V}}\hat{e}\cdot \nabla{}\left[\mathbf{\mu}(\mathbf{R})\right]\sum_{k=\mathrm{l,u}}\xi_{k}(c_k+c^\dagger_k)
\end{eqnarray}
\end{subequations}
where \(\omega_k\) is the frequency of each of the polariton mode, and \(\Gamma_k\) the decay. 
\(\bar{n}_{k}^{\mathrm{th}}\) is the mean
photon number of a mode \(\omega_k\) at temperature \(T\). \(\xi_{k}\) is the photon component
of the polariton mode. \(\alpha\) is the damping rate of the reactive subsystem. 
We assume that \(f_k\) and \(F_0\) are both white noises, i.e., their autocorrelations are $\delta$--functions
\( \langle f_k(t)f_{k'}(t')\rangle=\delta_{k,k'}\beta_k\delta(t-t') \) and
\( \langle F_0(t)F_0(t')\rangle=\delta(t-t') \).

\begin{figure}
    \centering
    \includegraphics[width=\linewidth]{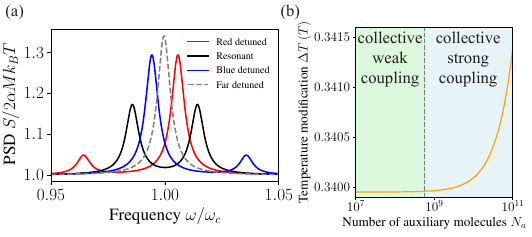}
    \caption{Properties of stochastic force \(F_{\mathrm{ST}}\). (a) Power spectrum density(PSD) 
    of \(F_{\mathrm{ST}}\) for different auxiliary ensemble detuning, normalized to the amplitude of the 
    white noise of a bare molecule. A single resonant peak corresponds to absence of auxiliary ensemble or a non-operational far-detuned ensemble. Tuning the emsemble to resonance splits the peak into two balanced components. By blue-detuning the ensemble, the principal noise peak could be red-detuned, and the primary peak moves to higher frequency when the ensemble is red-detuned. (b) Modification of effective temperature versus molecule numbers of the auxiliary ensemble. Increasing the molecule number moves the BTC subsystem into collective strong coupling regime, with a nonlinearly-growing effective temperature.}
    \label{fig:2}
\end{figure}

We formally solve Eqs.~(\ref{eq:eom_btc}) and by eliminating BTC variables, we obtain the modified equation of motion for the reactive subsystem in an optical cavity coupled to an auxiliary ensemble,
\begin{subequations}
\label{eq:eom_red}
\begin{equation}
\label{eq:eom_rea}
M\frac{\mathrm{d}^2 \mathbf{R}}{\mathrm{d} t^2}=
-\nabla \left[ E(\mathbf{R})\right]
-M\alpha{}\frac{\mathrm{d}\mathbf{R}}{\mathrm{d} t}+F_{\mathrm{FB}}
+F_{\mathrm{ST}}
\end{equation}
\begin{eqnarray}
F_{\mathrm{ST}}=&&-\sqrt{\frac{\hbar \omega_{\mathrm{c}}}{2\epsilon_0 V}}\hat{e}\cdot \nabla{}\left[\mathbf{\mu}(\mathbf{R})\right]\sum_{k=l,u}\xi_{k}(c_{k,\mathrm{ST}}+c^\dagger_{k,\mathrm{ST}})\nonumber\\
&&+\sqrt{2M\alpha k_{\mathrm{B}}T}F_0(t)
\end{eqnarray}
\begin{equation}
c_{k,\mathrm{ST}}(t)=\frac{1}{2\pi}\int_{-\infty}^\infty \frac{\sqrt{\Gamma_k \bar{n}^k_{th}}}{i(\omega_k-\omega)+\Gamma_k/2}f_k(\omega)e^{-i\omega t}\mathrm{d}\omega
\end{equation}
\begin{equation}
F_{\mathrm{FB}}=-\sqrt{\frac{\hbar \omega_{\mathrm{c}}}{2\epsilon_0 V}}\hat{e}\cdot \nabla{}\left[\mathbf{\mu}(\mathbf{R})\right]\sum_{k=l,u}\xi_{k}(c_{k,\mathrm{FB}}+c^\dagger_{k,\mathrm{FB}})
\end{equation}
\begin{equation}
c_{k,\mathrm{FB}}(t)=\frac{1}{2\pi}\int_{-\infty}^\infty \frac{ -\frac{i}{\hbar}\xi_{0,k}\sqrt{\frac{\hbar\omega_{\mathrm{c}}}{2\epsilon_0V}}\hat{e}\cdot\tilde{\mathbf{\mu}}(\mathbf{R}) }{i(\omega_k-\omega)+\Gamma_k/2}e^{-i\omega t}\mathrm{d}\omega
\end{equation}
\end{subequations}
Here \(\tilde{\mathbf{\mu}}(\mathbf{R})\) is the Fourier transform of \(\mathbf{\mu}[\mathbf{R}(t)]\).
In Eq.~\ref{eq:eom_rea}, \(F_{\mathrm{ST}}\) is the stochastic force on the reactive molecule,
containing the intrinsic noise and the noise induced by interaction.
\(F_{\mathrm{FB}}\) is the feedback as a backaction from the BTC subsystem,
from the electromagnetic field coherently perturbed by the molecular motion.
In the following, we examine these two forces more closely, highlighting their characteristics of noise and feedback.
For the stochastic force \(F_{\mathrm{ST}}\), coupling to the cavity results in a colored noise that originates from partially tracking out the white noise in the BTC subsystem.
\(F_{\mathrm{ST}}\) would be position dependent for a generic polarization,
and for the sake of simplicity we study the case of linear polarization, i.e.,
\( \chi=\sqrt{\frac{\hbar \omega_{\mathrm{c}}}{2\epsilon_0 V}}\hat{e}\cdot \nabla{}\left[\mathbf{\mu}(\mathbf{R})\right] \) is taken to be a constant.
We could obtain the power spectrum density (PSD) of the total noise in the stochastic force:
\begin{equation}
S_{F_{\mathrm{ST}}F_{\mathrm{ST}}}= 2\alpha M k_{\mathrm{B}}T+\chi^2\sum_{k=\mathrm{l,u}}
\frac{|\xi_{k}|^2\beta_k\bar{n}^{\mathrm{th}}_{k}\Gamma_k}{\Gamma_k^2/4+(\omega-\omega_k)^2}
\end{equation}
Its dependence on frequency represents a colored noise (Fig.~\ref{fig:2}(a)).
The line shapes are imprinted by the polariton modes, since the PSD structure essentially reflects the memory effect of the BTC subsystem. 
For a bare cavity, there is only a sole peak at the cavity mode.
Introducing a resonant and collectively strong-coupled ensemble splits the peak
into two balanced polariton branches.
By detuning the auxiliary ensemble, the peak could be shifted away from resonance with reactive molecules:
a red-detuned ensemble shifts the peak to a higher frequency, and a blue-detuned one shifts the peak to a lower frequency.

\begin{figure}
    \centering
    \includegraphics[width=\linewidth]{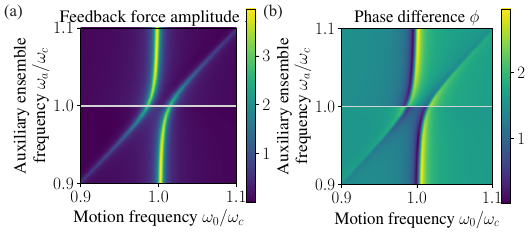}
    \caption{Properties of coherent feedback \(F_{\mathrm{FB}}\). 
    (a) Dependence of dimensionless amplitude \(A\) for trial frequency \(\omega_0\) on auxiliary ensemble detuning \(\Delta=\omega_{\mathrm{a}}-\omega_{\mathrm{c}}\). Light gray line indicates perfect resonance scenario. (b) Dependence of phase difference \(\phi\) for trial frequency \(\omega_0\) on auxiliary ensemble detuning \(\Delta=\omega_{\mathrm{a}}-\omega_{\mathrm{c}}\). Note that the phase difference is always positive, corresponding to negative feedback. Light gray line indicates the perfect resonance scenario. }
    \label{fig:3}
\end{figure}

When we ignore memory effect of the added noise,
then the cavity effect could be considered effectively as a modification of temperature~\cite{suppl}:
\begin{equation}
T_{\mathrm{eff}}=T+\frac{2\chi^2}{\alpha M k_{\mathrm{B}}}\sum_{k=\mathrm{l,u}}|\xi_{k}|^2\frac{\bar{n}^{\mathrm{th}}_{k}}{\Gamma_k}
\end{equation}
Fig.~\ref{fig:2}(b) shows the dependence of the effective temperature on the molecule number of the auxiliary ensemble. 
A nonlinear response is observed when the total number of thermal excitations in both polariton modes is varied.

\begin{figure}
    \centering
    \includegraphics[width=\linewidth]{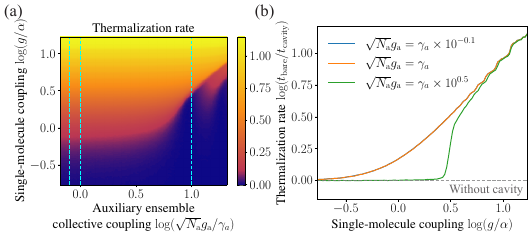}
    \caption{Simulation results of thermalization acceleration. (a)The thermalization time of intracavity molecules relative to that of bare molecules is shown. Improving single-molecule coupling strength generally allows faster decay. For a fixed single-molecule coupling, increasing the number of auxiliary molecules would erase the acceleration. Note that as coupling becomes larger, a sheerer transition from accelerated region to normal region is witnessed in increasing molecule numbers. (b)Dependence of thermalization rate on single-molecule coupling strength for fixed BTC background in collective weak or strong coupling regime. Note that the weak couple line is shaded by the \(\sqrt{N_a}=\gamma_a\) line, and the gray line indicates thermalization rate of bare molecules.}
    \label{fig:4}
\end{figure}

Now we proceed to the feedback force \(F_{\mathrm{FB}}\),
as a deterministic force for given temporal evolution \(R(t)\).
The real-time motion \(R(t)\) can be separated into components of different frequency
by applying a Fourier transform.
Each of these components individually generates a feedback of the same frequency,
with definite phase difference and amplitude.
In this spirit, we can always obtain the feedback for a specific motion by
calculating the feedback contributions from each of the motion components,
and sum them into the total feedback force \(F_{\mathrm{FB}}\).
For a mono-frequency motion component
\(R^{(0)}(t)=\sqrt{\frac{\hbar}{2M\omega_0}}\sin{\omega_0 t}\),
the corresponding feedback reads:
\begin{subequations}
\begin{equation}
F_{\mathrm{FB}}^{(0)}(t)=-\chi^2\sqrt{\frac{\hbar}{2M\omega_0}}\frac{1}{\hbar\alpha}A(\omega_0)\sin[\omega_0t+\phi(\omega_0)]
\end{equation} 
\begin{equation}
\phi=\arccos{}\left[\frac{\alpha\xi_{\mathrm{l}}(\omega_{\mathrm{l}}-\omega_0)}{(\omega_{\mathrm{l}}-\omega_0)^2+\Gamma_{\mathrm{l}}^2/4} +
\frac{\alpha\xi_{\mathrm{u}}(\omega_{\mathrm{u}}-\omega_0)}{(\omega_{\mathrm{u}}-\omega_0)^2+\Gamma_{\mathrm{u}}^2/4}\right]
\end{equation} 
\end{subequations}
It is a coherent feedback of the same angular frequency as the trial motion,
with dimensionless amplitude \(A\) and a phase shift \(\phi\) depending on \(\omega_0\).
Fig.~\ref{fig:3} shows dependence of \(A\) and \(\phi\) on \(\omega_0\) and system parameters.
The total feedback is a summation of backaction from both polariton modes.
It is shown that for such a system without pump or gain,
the phase shift \(\phi\) is always in the first or second quadrant
(See Supplemental Materials for explicit derivation for \(A\) and \(\phi\))~\cite{suppl}.
We then showcase the work done by the feedback force in one period, for \(\pi>\phi>0\) the work is negative,
which means the feedback from the cavity is always negative, namely, when the molecule is initially stimulated into non-equilibrium motion,
the feedback from the cavity will always do negative work and suppress molecule vibration,
which is a potential explanation to the lower fraction of high energy molecules 
observed in molecular dynamics simulations \cite{li2021collective}.

We have addressed that coupling to a cavity with an auxiliary molecule ensemble
introduces colored noise and negative feedback to reactive molecules.
While a negative feedback could aid decay of excited molecules to lower energies,
the ability of colored noise to accelerate thermalization has also been shown \cite{ceriotti2009langevin}.
Now we simulate the non-equilibrium dynamics of an oscillator,
initially prepared at a high excitation,
and track the time it takes to decay to half its initial energy as a metric for thermalization time (Fig. \ref{fig:4}).
It is noted that, while increasing single-molecule coupling strength unambiguously accelerates thermalization, the inclusion of more auxiliary molecules seems to cancel this boost.
As coupling strength becomes larger, the transition between accelerated region and normal region turns sharper as molecule number increases.
This sharp transition is reminiscent of the experimentally observed criticality when molecule number is changed.

To conclude, we have examined the non-equilibrium stochastic processes in an open system of reactive molecules situated within an optical resonator absent of external stimuli. Our findings reveal that coupling introduces additional colored noise and coherent feedback, with the locations and magnitudes of peaks being adjustable by a supplementary ensemble. The added noise creates a non-Markovian setting for the reactive subsystem, which has been demonstrated to expedite thermalization \cite{ceriotti2009langevin}.
We have also shown that the coherent feedback is always negative and
thus inhibits molecule vibration, suppressing the high energy motion. 
Furthermore, we numerically simulated the non-equilibrium dynamics of an oscillator initially prepared at high excitation, from which the accelerated thermalization is quantitatively investigated.

\begin{acknowledgments}
This work was supported by National Key R\&D Plan of China (Grant No. 2023YFB2806702, 2023YFA1406801), National Natural Science Foundation of China (Grant Nos. 12293051, 32450712, 12474371, 12174009, 12234002, 92250303), the Beijing Natural Science Foundation (Grant No. Z220008), and the High-performance Computing Platform of Peking University.
\end{acknowledgments}

\bibliography{biblio}

\end{document}